\documentclass[a4paper,fleqn,usenatbib]{mnras}

\usepackage[T1]{fontenc}
\usepackage{ae,aecompl}

\usepackage{graphicx}	
\usepackage{amsmath}	
\usepackage{amssymb}	

\title[Studying topological structure of 21cm signal with 3D-MFs]{Studying topological structure of 21cm line fluctuations with 3D-Minkowski functionals before the reionization}
\author[S. Yoshiura et al.]{
 Shintaro Yoshiura$^1$\thanks{E-mail: 161d9002@st.kumamoto-u.ac.jp},
 Hayato Shimabukuro$^{1,2}$,
 Keitaro Takahashi$^1$,
 \newauthor
and Takahiko Matsubara$^{2,3}$
\\
$^{1}$Department of Physics, Kumamoto University, Kumamoto 860-8555,Japan\\
$^{2}$Department of Physics, Nagoya University, Chikusa, Nagoya 464-8602, Japan\\
$^{3}$Kobayashi-Maskawa Institute for the Origin of Particles and the Universe, Nagoya University, Chikusa, Nagoya 464-8602, Japan
}

\date{Accepted XXX. Received YYY; in original form ZZZ}

\pubyear{2016}

\begin{document}
\label{firstpage}
\pagerange{\pageref{firstpage}--\pageref{lastpage}}
\maketitle

\begin{abstract}
The brightness temperature of the redshifted 21cm line brings rich information on the Inter Galactic Medium (IGM) from the Cosmic Dawn and Epoch of Reionization (EoR). While the power spectrum is a useful tool to statistically investigate the 21cm signal, the 21cm brightness
temperature field is highly non-Gaussian, and the power spectrum is
inadequate to characterize the non-Gaussianity. The Minkowski Functionals (MFs) are promising tools to extract non-gaussian features of the 21cm signal and give topological information such as morphology of ionized bubbles. In this work, we study the 21cm line signal in detail with MFs. To promote understanding of basic features of the 21cm signal, we calculate the MFs of not only the hydrogen neutral fraction but the matter density and spin temperature, which contribute to the brightness temperature fluctuations. We find that the structure of the brightness temperature depends mainly on the ionized fraction and the spin temperature at late and early stages of the EoR, respectively. Further, we investigate the redshift evolution of the MFs at $7 < z < 20$. Then, we find that, after the onset of reionization, the MFs reflect mainly the ionized bubble property. In addition, the MFs are sensitive to model parameters which are related to the topology of ionized bubbles and we consider the possibility of constraining the parameters by the future 21cm signal observations.   
\end{abstract}




\section{Introduction}

After the cosmic recombination, hydrogen keeps the neutral state until first luminous objects such as stars and galaxies are formed.  As the structure formation advances, neutral hydrogen begins to be ionized by the UV and X-ray photons escaped from galaxies. This epoch is called ``Epoch of Reionization''. Many theoretical works based on both numerical simulations and analytic models have investigated the time evolution and spatial structure of ionized bubbles in the EoR \citep{2011MNRAS.411..955M,2007MNRAS.377.1043M,2007ApJ...654...12Z}. We can obtain the information on the physical state of the IGM in the EoR through the redshifted 21cm line due to the hyperfine structure of neutral hydrogen \citep{2006PhR...433..181F}.

However, the 21cm signal from the EoR has not been accessible yet. Ongoing interferometers such as the Murchison Widefield Array (MWA)\citep{2009IEEEP..97.1497L,2013PASA...30....7T,2013MNRAS.429L...5B}, the LOw Frequency ARray (LOFAR) \citep{2013A&A...556A...2V,2013MNRAS.435..460J}, and the Precision Array for Probing the Epoch of Reionization (PAPER) \citep{2010AJ....139.1468P,2014ApJ...788..106P} have already started observations to detect the 21cm signal statistically. In fact, the MWA and the PAPER have obtained upper limits on the 21cm power spectrum (e.g. \cite{2015PhRvD..91l3011D}). The Square Kilometre Array low (SKA-low), a future project of a low-frequency radio interferometer, will start observations in the near future \citep{2015aska.confE.171C} and it will have enough sensitivity to image the 21cm-line signal at scales from several arcminutes to several degrees \citep{2015aska.confE..10M,2015aska.confE..15W}.

In order to understand the thermal and ionized state of the IGM, the power spectrum has often been considered for the statistical analyses of the 21cm brightness temperature fluctuations. For example, the typical bubble size could be probed through the ``shoulder'' of the 21cm power spectrum \citep{2007MNRAS.376.1680P,2014ApJ...782...66P,shi-P,2006PhR...433..181F,2014MNRAS.439.3262M}. Although the power spectrum is a powerful tool, the statistical properties of
the 21cm-signal fluctuations are not perfectly described by the power
spectrum, because they follow a highly non-gaussian distribution due to complicated ionizing processes. Therefore, higher order statistics such as the bispectrum \citep{2016MNRAS.tmp..264S} and trispectrum \citep{2008PhRvD..77j3506C} are required to evaluate the non-gaussianity. 

The Minkowski Functionals (MFs) are another tool to study the non-gaussianity. As the reionization process advances, many ionized bubbles appear and these bubbles have typical topological structures in the brightness temperature field. The MFs are known to be useful to characterize this kind of topology. In the context of observational cosmology, the MFs have been used to investigate the geometrical feature of galaxy distribution \citep{1986ApJ...306..341G,1997ApJ...482L...1S} and to estimate the non-gaussianity appeared in the CMB temperature map \citep{2009ApJS..180..330K}.

There have been several works to study the MFs and genus to characterize the topology of HI distribution in the EoR. The topology varies with the evolution of ionization and the different stages of reionization could be distinguished by the MFs and genus \citep{2008ApJ...675....8L}. The topological property was utilized for suggesting a semi-numerical model which reproduces the MFs obtained from simulation with radiative transfer of ionizing photons \citep{2006MNRAS.370.1329G}, and for investigating dependency of HI topology on ionizing source property \citep{2011MNRAS.413.1353F}. In addition, the distribution of the 21cm line brightness temperature has been studied by the 2D genus \citep{2014JKAS...47...49H,2015ApJ...814....6W} . Then, the $\delta T_{\rm b}$ have been calculated with an assumption $T_{\rm S}\gg T_{\gamma}$ which is a suitable condition at the epoch where the spin temperature is sufficiently larger than the CMB temperature. They have found that genus curves depend on the ionized fraction, discriminate the scenarios with different ionizing efficiency values. 

In this paper, we calculate the 3-dimensional MFs of the 21cm brightness temperature. The brightness temperature fluctuations consist of three components: fluctuations in the neutral fraction, spin temperature and matter density. In the previous works, the MFs were calculated with an assumption that the spin-temperature fluctuations are negligible since the mean spin temperature is expected to be much larger than the CMB temperature as a result of X-ray heating. Although this assumption is frequently used, it depends on the redshift and the property of X-ray sources. Thus, first, we study the MFs of the three components and obtain an insight into the brightness temperature fluctuations. Secondly, we investigate the redshift evolution of the MFs at the epoch from cosmic dawn through reionization. Finally, to study the dependence of MFs on ionized bubble properties such as the size, shape and distribution, we calculate the MFs varying key model parameters. 

This paper is organized as follows. In section 2, we explain the brightness temperature, semi-numerical simulation 21cmFAST and our models. In section 3, we introduce the Minkowski Functionals. Our main results are presented in section 4. The summary and discussion will be given in section 5. Throughout this paper, we assume $\Lambda {\rm CDM}$  cosmology with ($\Omega_{\rm m}, \Omega_{\Lambda}, \Omega_{\rm b}, H_0, n_s, \sigma_8$) = ({0.31, 0.69, 0.048, 68km/s/Mpc, 0.97, 0.82) \citep{Planck2015}.}

\section{Set up : Simulation Model}

The observable quantity for the redshifted 21cm-line signal is the brightness temperature defined by the difference between the spin temperature and the CMB temperature \citep{2011MNRAS.411..955M,2006PhR...433..181F}, and can be written approximately as,
\begin{eqnarray}
\delta T_{\rm b}(z)
\approx 27 x_{\rm HI} (1 + \delta_{\rm m})
          \bigg(\frac{H}{dv_r/dr + H} \bigg)
          \bigg(1 - \frac{T_{\gamma}}{T_{\rm S}}\bigg)\nonumber\\
          \times\bigg(\frac{1+z}{10} \frac{0.15}{\Omega_{\rm m} h^2}\bigg)^{1/2}
          \bigg(\frac{\Omega_{\rm b} h^2}{0.023} \bigg) ~ [{\rm mK}],
\label{eq:brightness}
\end{eqnarray}
where $H$ is the Hubble parameter, $dv_r/dr$ is the peculiar velocity, $\Omega_{\rm m}$ is the matter density parameter, $\Omega_{\rm b}$ is the baryon density parameter, $h$ is the Hubble constant, $x_{\rm HI}$ is the neutral fraction, $\delta_{\rm m}$ is the density contrast of matter, $T_{{\rm S}}$ is the spin temperature and $T_{\gamma}$ is the CMB temperature. The spin temperature evolves through coupling to the kinetic temperature of the IGM by the Wouthuysen-Field (WF) effect \citep{2006MNRAS.367.1057P} and the CMB temperature. When the IGM is heated sufficiently and the spin temperature becomes much higher than the CMB temperature, the factor (1-$T_{\gamma}$/$T_{\rm S}$) is saturated to become unity. Therefore, the $T_{\rm S}$ is often ignored at the late phase of the EoR ($z \lesssim 10$). The fluctuations in $\delta T_{\rm b}$ are expressed by
\begin{equation}
\delta_{21}({\bf x},z) =\frac{ \delta T_{\rm b}({\bf x},z) - \overline{\delta T_{\rm b}}(z)}{|\overline{\delta T_{\rm b}}(z)|}.
\label{delta_T_b}
\end{equation}
Here, $\overline{\delta T_{\rm b}}$ is the spacial average of the brightness temperature.

We make brightness temperature maps by a public semi-numerical code 21cmFAST \citep{2011MNRAS.411..955M}, setting a $400^3\rm Mpc^3$ comoving box with $600^3$ cells. From the simulations, not only brightness temperature maps but those of the neutral fraction $x_{\rm HI}$, matter density $\delta_{\rm m}$ and spin temperature $T_{\rm S}$ can be obtained at each redshift. Although the 21cmFAST adopts some simplifying models for the calculations of the matter density distribution and radiative transfer \citep{2004ApJ...613....1F}, the resulting maps are reasonably consistent with those from more sophisticated numerical simulations at large scales ($\gtrsim O(1)~{\rm Mpc}$) \citep{2011MNRAS.411..955M}.

\begin{figure}
 \centering
 \includegraphics[width=8cm]{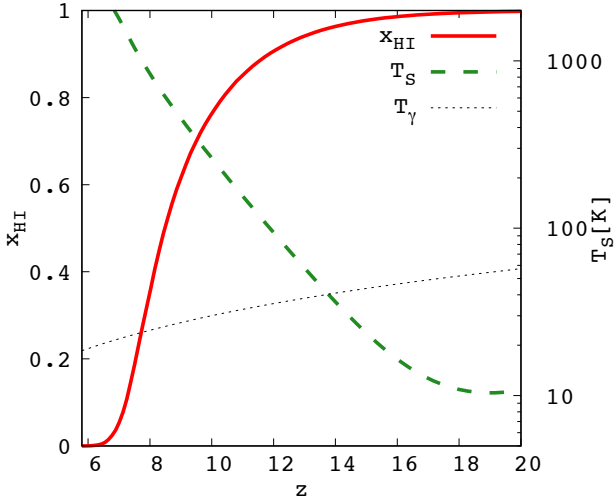}
  \caption{Evolution of $x_{\rm HI}$ and $T_{\rm S}$ for our ``fid'' model. The CMB temperature $T_{\gamma}$ is also shown for reference. } 
 \label{xHTs} %
\end{figure}

The 21cmFAST has three key parameters: (1) the ionizing efficiency ($\zeta_{\rm ion}$) which is a product of the fraction of gas converted into stars, the escape fraction and number of ionizing photon produced per stellar baryon \citep{2014MNRAS.439.3262M}, (2) the minimum virial temperature of halos which host galaxies ($T_{\rm vir}^{\rm min}$) and (3) the number of X-ray photons emitted from sources per solar mass ($\zeta_X$). We take ($\zeta_{\rm ion}, T_{\rm vir}^{\rm min},\zeta_X$) = ($15.0, 1\times10^4, 10^{56}$) as our fiducial model (hereafter ``fid'' model), with which the reionization is completed at $z \approx 6$ as indicated from observations of high-$z$ quasar spectra \citep{2006AJ....132..117F}. The value of $T_{\rm vir}^{\rm min} = 1 \times 10^4[{\rm K}]$ corresponds to the criterion of the effectiveness of the atomic cooling. Below this temperature, the atomic cooling is ineffective and the molecular cooling due to the oscillation and rotation of molecular hydrogens becomes the main cooling process. However, if radiative feedback processes such as photodissociation are effective, molecular hydrogens are destroyed and star formation in halos is suppressed \citep{2000ApJ...534...11H,2013MNRAS.428..154H}. Thus, $T_{\rm vir}^{\rm min}$ parametrizes the effectiveness of radiative feedback processes. The evolution of the neutral fraction ($x_{\rm HI}$) and the spin temperature for the ``fid'' model is shown in Fig.~\ref{xHTs}.

To understand the MFs of the brightness temperature, we prepare three kinds of brightness temperature maps.
\begin{itemize}
\item Component map ``$\rm T_{\rm S}$'': only the spin temperature fluctuations are taken into account, ignoring the fluctuations in $x_{\rm HI}$ and $\delta_{\rm m}$ (i.e. $x_{\rm HI} ({\bf x},z)=\overline{x_{\rm HI}}(z)$ and $\delta_{\rm m}({\bf x})=0$).
\item Component map ``m'':  only the matter density fluctuations are taken into account, while we take the average value of $x_{\rm HI}$ and set $T_{\gamma}/T_{\rm S}({\bf x}) = 0$.
\item Component map ``$\rm x_{\rm H}$'': only the fluctuations in the neutral hydrogen fraction are taken into account and we set $\delta_m(\bf{x})=0$ and $T_{\gamma}/T_{\rm S}({\bf x})=0$.
\end{itemize}
The MFs calculated from these three maps represent the structure of the components which contribute to the brightness temperature fluctuations. In section \ref{subsec:Comp}, we compare the MFs of these maps with that of the full brightness temperature.

Below, we will see the dependence of the MFs on the model parameters. Here we focus on two parameters, $\zeta_{\rm ion}$ and $T_{\rm vir}^{\rm min}$ because the topology of ionized bubbles is expected to be affected significantly by these parameters \citep{2013MNRAS.428.2467K}.

\section{Minkowski Functionals}

We can extract topological information on an arbitrary scalar function $p$(${\bf{x}}$) distributed in 3D space via the MFs. The excursion set $F_{\nu}$ consists of all $p({\bf{x}}$) which satisfy $p({\bf{x}}) > p_{{\rm th}}$. Here, $p_{\rm th}$ is the threshold value defined as $p_{\rm th} = \nu \sigma$ where $\sigma^2 (= \langle p^2({\bf{x}}) \rangle$) is the variance of $p({\bf{x}})$ while the average is zero in the volume $V$. The definitions of MFs are given by \citep{1997ApJ...482L...1S, 2006MNRAS.370.1329G} as a function of $\nu$,
\begin{eqnarray}
V_0(\nu)&=&\frac{1}{V}\int_V d^3x \Theta (p({\bf{x}})-\nu \sigma),\\
V_1(\nu)&=&\frac{1}{6V}\int_{\partial F_{\nu}}ds,\\
V_2(\nu)&=&\frac{1}{6\pi V}\int_{\partial F_{\nu}}ds[\kappa_1({\bf{x}})+\kappa_2({\bf{x}})],\\
V_3(\nu)&=&\frac{1}{4\pi V}\int_{\partial F_{\nu}}ds\kappa_1({\bf{x}})\kappa_2({\bf{x}}),
\end{eqnarray}
where $\Theta$ is the Heaviside step function, $\partial F_{\nu}$ is the surface of excursion set and $ds$ is the surface element. $\rm \kappa_1$ and $\kappa_2$ are curvatures on the surface. The $V_0$ represents the volume of excursion set. The total surface and mean curvature of $F_{\nu}$ are given by $V_1$ and $V_2$, respectively. The $V_3$ is the Euler characteristic which characterizes the typical shape of $F_\nu$ and is related to the genus $g$ as $V_3=2(1-g)$. For instance, a sphere has $V_3 = 2$ ($g=0$) and a torus has $V_3 = 0$ ($g=1$). In general, when the $V_3$ is positive, there are a lot of isolated excursion sets. Negative value of $V_3$, on the other hand, indicates that the excursion set has a multi-connected structure. The calculation method of the MFs is shown in \citet{2006MNRAS.370.1329G} and \citet{1997ApJ...482L...1S}. 

\begin{figure}
 \centering
  \includegraphics[width=7cm,clip]{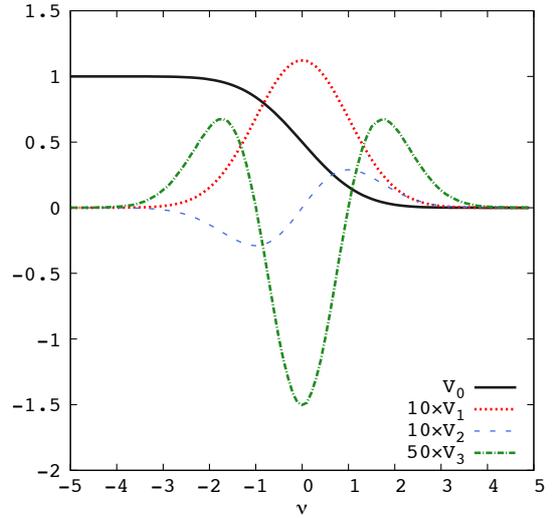}
   \caption{The Minkowski fuctionals for gaussian distribution. } 
 \label{fig:gaus} %
\end{figure}

When $p(x)$ has a gaussian distribution, the MFs can be written analytically and have symmetric shapes, as shown in Fig.~\ref{fig:gaus}. We can see that the $V_3$ has two positive peaks and a valley at center of $\nu$ which corresponds to the average value of $p(x)$. These two peaks reflect the symmetric probability distribution function of the gaussian distribution with the same number of isolated regions which have positive and negative values. For a non-gaussian distribution, this symmetry is broken. When the non-gaussinaity is weak, the MFs can be described analytically \citep{1994ApJ...434L..43M,2003ApJ...584....1M}. As we will see below, the MFs of the $\delta T_{\rm b}$ are significantly different from those of the gaussian case because the $\delta T_{\rm b}$ follows a highly non-gaussian distribution.

We use $p(x)$ of the brightness temperature maps smoothed by a gaussian kernel which is expressed by $\propto \exp(-x^2/2R^2)$ where $R$ is a smoothing scale. Throughout this paper, we set $R = 3~{\rm Mpc}$. Actually, the shapes of MFs are dependent on the smoothing scale, but we found the dependence is not so large for $R > 3~{\rm Mpc}$. The angular resolution of the SKA1-low will be $\rm FWHM \sim 5~arcmin$, which corresponds to $R = 5.8~{\rm Mpc}$, at $140~{\rm MHz}~(z=9)$. In fact, the FWHM depends on the frequency (redshift). For example, the resolution at which the SKA1-low has the best sensitivity becomes twice for $z=19$. Note that an ionized bubble with a radius smaller than the smoothing scale $R$ is erased by the smoothing effect. Therefore, although the brightness temperature at ionized regions should be zero, this does not become precisely zero if we perform the smoothing with a smoothing scale larger than them. We call this region ``unresolved ionized'' region.

\section{Results}

The 21cm brightness temperature is expected to follow a non-gaussian distribution due to highly-nonlinear astrophysical effects. The MFs are useful to evaluate complicated topological structure of this kind of 3D map. However, it is not straightforward to obtain a physical interpretation from the MFs. Therefore, in this section to promote understanding the MFs, we try to comprehend topological features of the brightness temperature distribution through three subsections.

\begin{figure*}
 \centering
  \includegraphics[width=15cm]{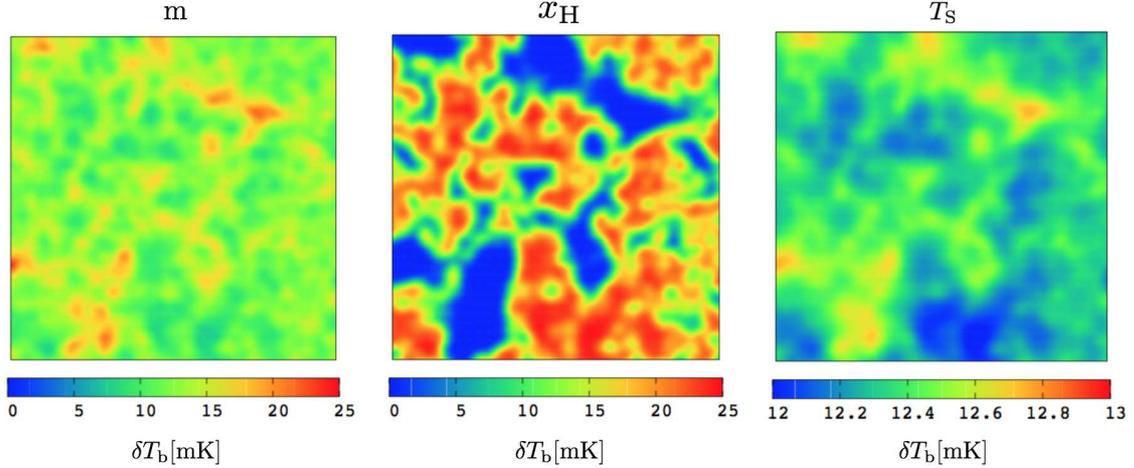}
   \caption{Maps of the brightness temperature for three components at $z = 8.62$. The panels are component maps ``m'', ``$\rm x_H$'' and  ``$\rm T_S$''  from left to right. Each slice is 200 Mpc on a side and 0.667 Mpc thick. Note the difference in the color bar scale for the right panel.} 
 \label{fig:comp} %
\end{figure*}

\begin{figure*}
 \centering
 \includegraphics[width=11cm]{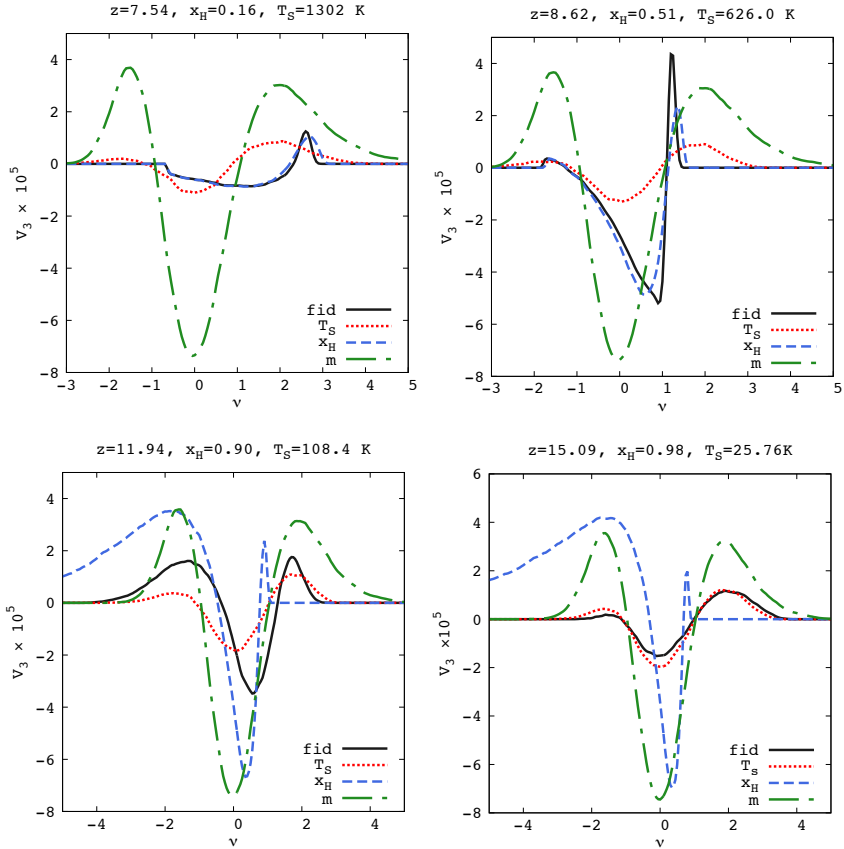}
  \caption{Comparison of a MF $V_3$ among ``fid'' (solid line), ``$\rm T_S$'' (dotted line), ``$\rm x_H$'' (dashed line) and ``m'' (dot dashed line) models. The redshifts are 7.54, 8.62, 11.94 and 15.09 from top to bottom.} 
 \label{fig:MFs_z} %
\end{figure*}

\begin{figure*}
 \centering
  \includegraphics[width=13cm,clip]{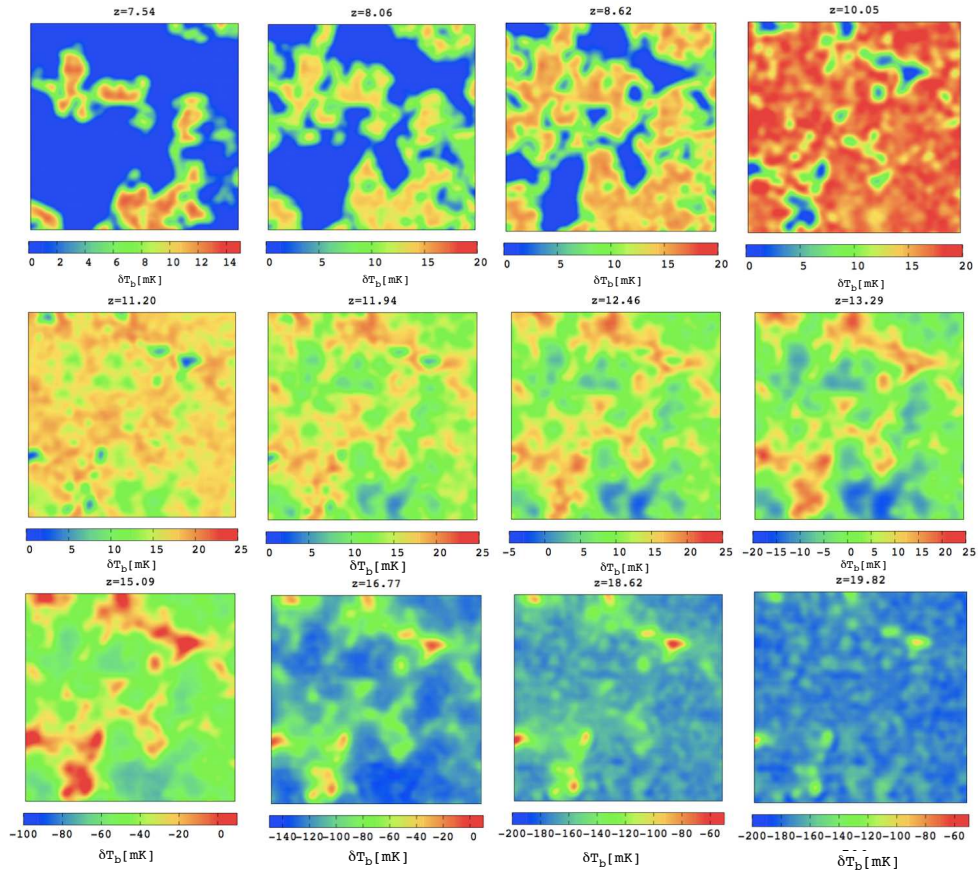}
   \caption{Redshift evolution of brightness temperature maps of the ``fid''  model for redshifts $z$=7.54 $\sim$ 19.82. Each slice is 200 Mpc on a side and 0.667 Mpc thick.} 
 \label{fig:map_def} %
\end{figure*}

\begin{table}
\centering
  \begin{tabular}{|l|c|r|} \hline
    map ($z$=7.54) & $\overline{\delta T_{\rm b}}$ &$\sigma_{\delta_{21}}$ \\ \hline
    fid & 2.54 & 1.55  \\
    $x_{\rm H}$ & 3.90 & 1.57  \\
    $T_{\rm S}$  & 3.83 & 0.00296 \\
    $m$  & 3.90 & 0.180  \\ \hline
  \end{tabular}
  \begin{tabular}{|l|c|r|} \hline
    map ($z$=8.62)& $\overline{\delta T_{\rm b}}$ &$ \sigma_{\delta_{21}}$ \\ \hline
    fid & 9.45 &  0.559 \\
    $x_{\rm H}$ & 12.9 & 0.582 \\
    $T_{\rm S}$  & 12.3 & 0.0114 \\
    $m$  & 12.9 & 0.160  \\ \hline
  \end{tabular}
  \begin{tabular}{|l|c|r|} \hline
    map ($z$=11.94)& $\overline{\delta T_{\rm b}}$ & $\sigma_{\delta_{21}}$ \\ \hline
    fid & 14.1 & 0.219  \\
    $x_{\rm H}$ & 26.6 & 0.107 \\
    $T_{\rm S}$  & 15.0 & 0.232 \\
    $m$  & 26.6 & 0.119  \\ \hline
  \end{tabular}
  \begin{tabular}{|l|c|r|} \hline
    map ($z$=15.09)& $\overline{\delta T_{\rm b}}$ & $\sigma_{\delta_{21}}$ \\ \hline
    fid & -42.1 & 0.411  \\
    $x_{\rm H}$ & 32.3 & 0.0273 \\
    $T_{\rm S}$  & -44.2 & 0.451 \\
    $m$  & 32.3 & 0.0955  \\ \hline
  \end{tabular}
  \caption{The average value of $\delta T_{\rm b}$ and standard deviation of $\delta_{21}$ for the ``fid'' and component maps at $z$=7.54, 8.62, 11.94 and 15.09. The threshold value is calculated from $\sigma$ as $p_{th}=\nu \sigma_{\delta_{21}}$. The $p_{\rm th}$ is converted into $\delta T_{\rm b} = \overline{\delta T_{\rm b}}+p_{\rm th}| \overline{\delta T_{\rm b}}|$.}
  \label{table}
\end{table}

\subsection{Components of brightness temperature \label{subsec:Comp}}

The topology of the brightness temperature in the EoR is expected to reflect the three components, the spin temperature, neutral hydrogen fraction and matter density. In this subsection, we compare the $V_3$ of the ``fid'' model with that of ``$\rm T_{\rm S}$'', ``m'' and ``$\rm x_{\rm H}$'' models to study their relative importance in the structure of $\delta T_{\rm b}$.

In Fig.~\ref{fig:comp}, we show the brightness temperature maps for the three components at $z=8.62$. The average value of the spin temperature at this redshift is $626.0~[{\rm K}]$ which is much larger than the CMB temperature $T_{\gamma}(z=8.62) = 26.12~[{\rm K}]$. In this case, the map of ``$\rm T_{\rm S}$'' has much smaller fluctuations compared with the other components and the spin temperature fluctuations do not contribute to $\delta T_{\rm b}$. In the middle panel, we can see a lot of evolved ionized bubbles located in dense regions and have $\delta T_{\rm b} = 0$.

Fig.~\ref{fig:MFs_z} shows the $V_3$ of the four models, ``fid'', ``$\rm T_{{\rm S}}$'', ``m'' and ``$\rm x_{{\rm H}}$'' at redshifts $z$ = 7.54, 8.62, 11.94 and 15.09. 
For reference, in table \ref{table}, we list the value of $\overline{\delta T_{\rm b}}$ and $\sigma_{\delta_{21}}$ of the models at the redshifts. 
In addition, in Fig.~\ref{fig:map_def}, we show $\delta T_{\rm b}$ maps of the ``fid'' model at several redshifts. 
Below, we discuss both the behavior of the $V_3$ for ``fid'' model and the relation between the ``fid'' and component maps.

At $z = 7.54$, as one can see in the top left panel of Fig.~\ref{fig:MFs_z}, the $V_3$ of ``m'' has two positive peaks and one valley similar to the case of a gaussian distribution, while the symmetry is broken. This is considered to be due to the nonlinearity of gravity. On the other hand, ``$\rm x_{\rm H}$'' and ``fid'' models have similar shapes. This suggests that, at this redshift, the topology of $\delta T_{\rm b}$ is mostly determined by the distribution of ionized fraction, while the spin temperature and matter density fluctuations have little effects. Further, the positive peak at a lower $\nu$ cannot be seen for ``$\rm x_{\rm H}$'' and ``fid'' models. This can be understood as follows. The $\delta T_{\rm b}$ is close to zero in dense regions where the IGM was already ionized. Because the positive peak at lower $\nu$ represents the abundance of isolated small-$\delta T_{\rm b}$ region, this peak disappears after ionized bubbles merged as can be seen in Fig.~\ref{fig:map_def}.

Next, the $V_3$ of the ``fid'' model at $z = 8.62$ is similar to that at $z = 7.54$ and still traces mostly that of ``$\rm x_{\rm H}$'' model. However, it has a small positive peak at a lower $\nu$ ($\nu \approx -1.5$). As we can see in Fig.~\ref{fig:map_def}, some ionized bubbles are left at this redshift and they contribute to this peak. 

At $z = 11.94$, the average spin temperature is about $108.4~[{\rm K}]$, which is not much higher than the CMB temperature $T_\gamma = 36~[{\rm K}]$. In this epoch, neutral hydrogen begins to be ionized at dense regions. The shape of the $V_3$ does not trace any single component as we can see in the bottom left panel of Fig.~\ref{fig:MFs_z}. Therefore, it is expected that all components contribute to the structure of the brightness temperature.

As we can see in the bottom right panel of Fig.~\ref{fig:MFs_z}, the $V_3$ of ``fid'' model is similar to that of the ``$T_{\rm S}$'' model at $z = 15.09$. At this redshift, most of hydrogen are neutral and the $V_3$ indicates that the main component of $\delta T_{\rm b}$ is ``$T_{\rm S}$''. 

In the literature, an assumption $T_{\rm S} \gg T_{\gamma}$ is often adopted at low redshifts to neglect the spin temperature fluctuations because this significantly reduces the computational cost. In our model, the average spin temperature is three times higher than $T_{\gamma}$ at $z = 11.94$ and one may consider this assumption is valid. However, the $V_3$ at this redshift has an asymmetry due to the spin temperature fluctuations. Thus, even though $T_{\rm S} \gg T_{\gamma}$ seems reasonable for the average spin temperature, we should be careful when we consider the MFs.

\begin{figure*}
 \centering
 \includegraphics[width=11cm]{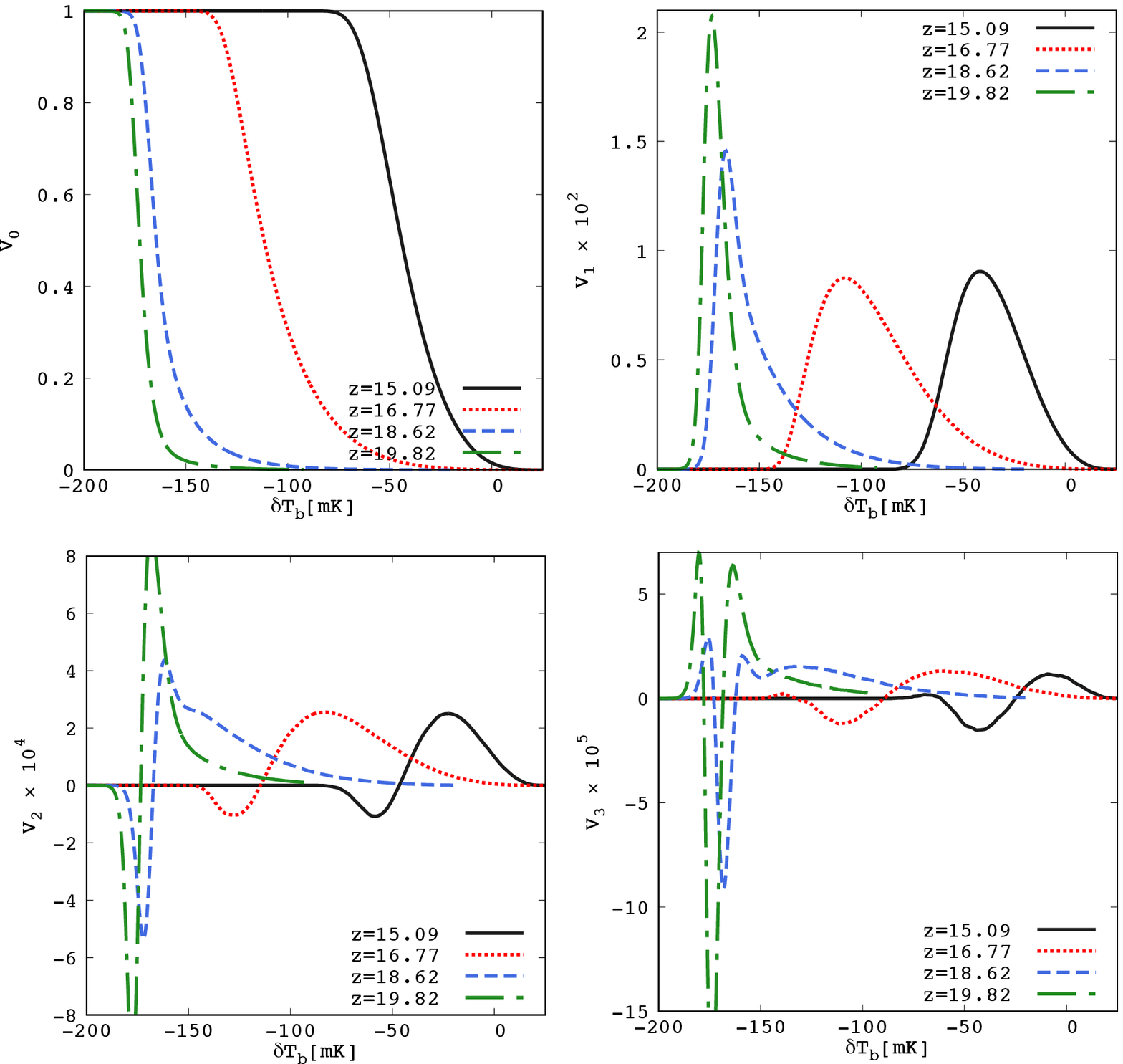}
  \caption{MFs of the brightness temperature of the ``fid'' model at the heating stage, $z = 15.09, 16.77, 18.62$ and 19.82. } 
 \label{evolution3} %
\end{figure*}

\begin{figure*}
 \centering
 \includegraphics[width=11cm]{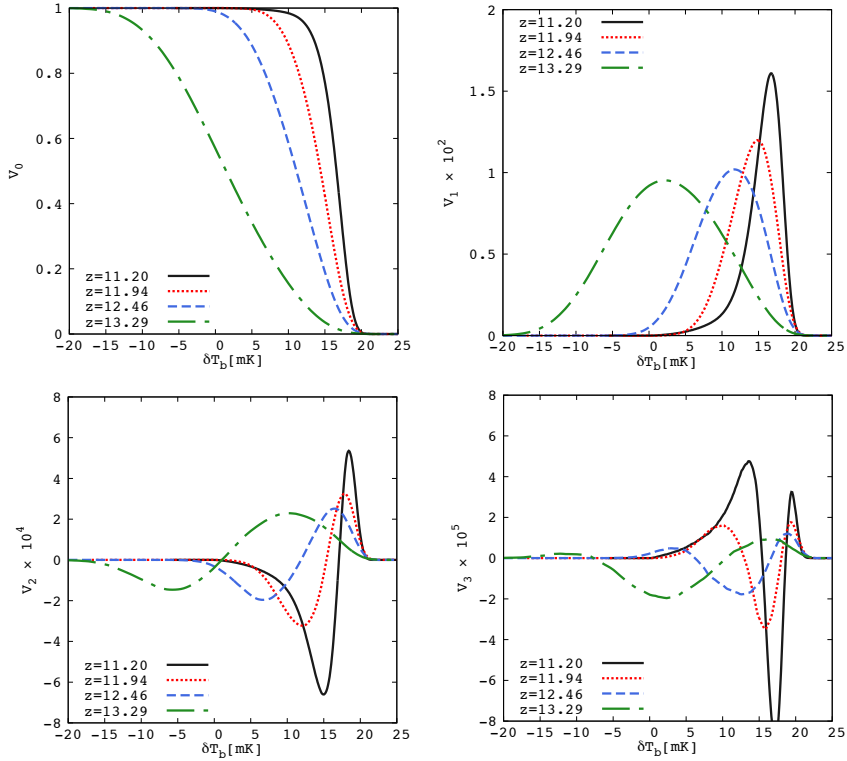}
  \caption{MFs of the brightness temperature of the ``fid'' model at the transition stage, $z=11.20, 11.94, 12.46$ and 13.29. } 
 \label{evolution2} %
\end{figure*}

\begin{figure*}
 \centering
 \includegraphics[width=11cm]{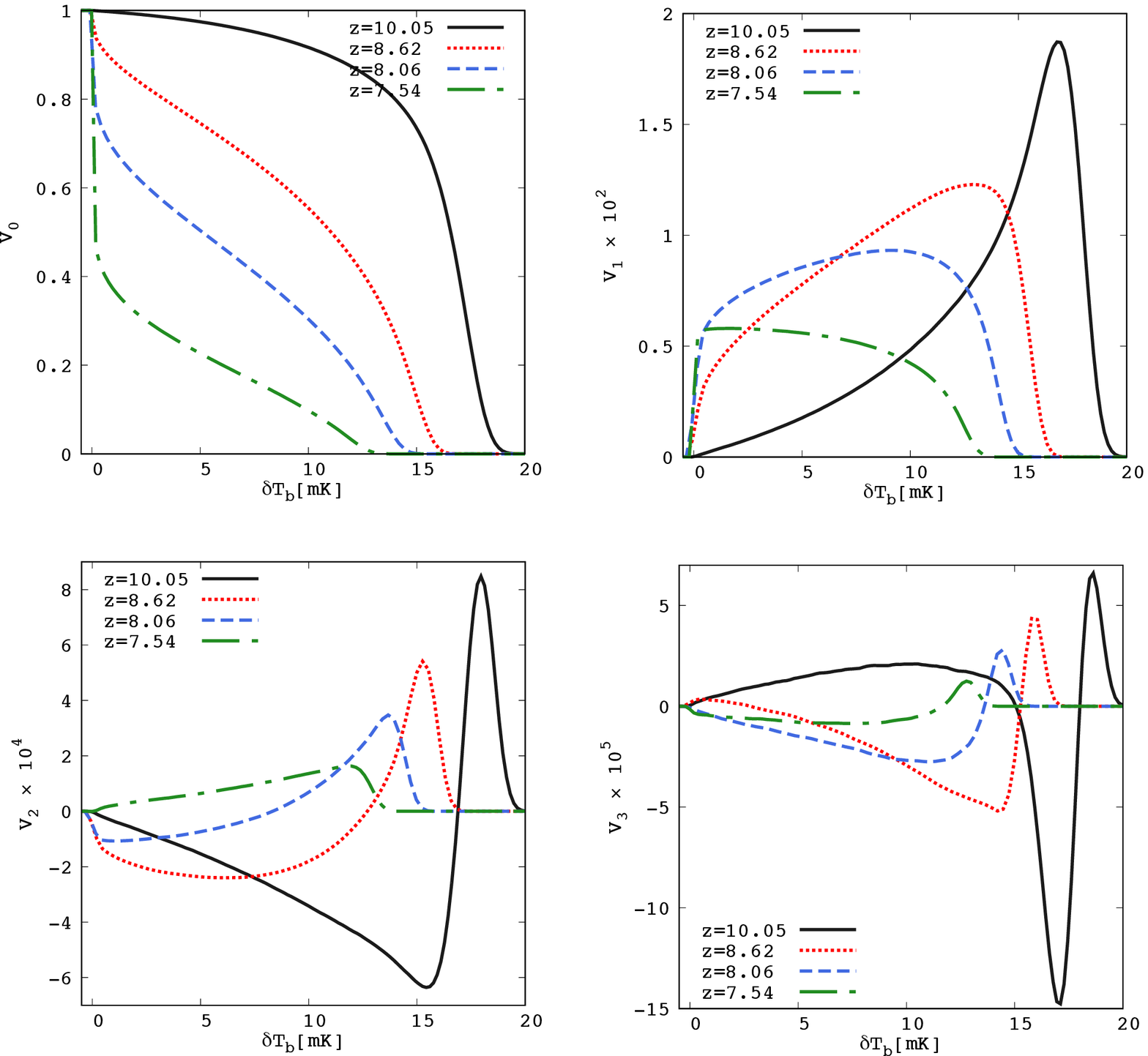}
  \caption{MFs of the brightness temperature of the ``fid'' model at the reionization stage, $z=7.54, 8.06, 8.62$ and 10.05. } 
 \label{evolution1} %
\end{figure*}

\subsection{Redshift evolution of MFs}

Here, we focus on the redshift evolution of MFs. The Figs.~\ref{evolution3} - \ref{evolution1} show the MFs as a function of $\delta T_{\rm b}$, instead of $\nu$, at twelve redshifts from 7.54 to 19.82. In Fig.~\ref{fig:map_def}, we show the maps of $\delta T_{\rm b}$ at these redshifts for reference. This period can be divided into three parts: (1) the heating stage ($z \gtrsim 15.09$) where the main component of $\delta T_{\rm b}$ is $T_{\rm S}$, (2) the transition stage ($11.20 < z < 13.29$) where the main component varies from $T_{\rm S}$ to $x_{\rm HI}$ and (3) the reionization stage ($z < 10.05$) where the main component is $x_{\rm HI}$. Before discussing the redshift evolution, we note that almost all of the MFs at Fig.~\ref{evolution3}- \ref{evolution1} are clearly different from the MFs for a gaussian field. The asymmetric shape of the MFs indicates the $\delta T_{\rm b}$ follows a non-gaussian distribution at any redshift.

First, let us consider the heating stage. In Fig.~\ref{evolution3}, the MFs in this stage are shown. 
As X-ray heating progresses, the MFs shift from low to high values of $\delta T_{\rm b}$ and change their shapes. However, the shape of $V_0$ is similar to each other, while the $V_1$ changes its shape drastically. At $z = 19.82$, the $V_1$ has a long tail on the right. This tail can be attributed to rare small heated regions. As heated regions expand and become more abundant, the $V_1$ approaches to a symmetric shape similar to the gaussian one. The $V_2$ also has this tendency. Concerning the $V_3$, the peak on the right has a long tail which indicates that small heated regions are isolated. Another peak on the left represents the existence of isolated cold regions. However, this peak becomes smaller as heating progresses. This is because X-ray photons have a long mean free path and the heating effect is not limited to the neighborhood of the sources. 

Second, let us see the transition stage. In Fig.~\ref{evolution2}, we show the MFs in this stage. Although the average value of $\delta T_{\rm b}$ is positive at this stage, a small fraction of pixels has a negative $\delta T_{\rm b}$ ($T_{\rm S} < T_{\gamma}$) at $z = 13.29$ and 12.46, while $\delta T_{\rm b}$ becomes positive at all pixels at later redshifts. At $z = 13.29$, the $V_0$ is 0.5 at $\delta T_{\rm b} = 0$, which means that the number of hot regions ($T_{\rm S} > {T_{\gamma}}$) and cold regions ($T_{\rm S}>{T_{\gamma}}$) are almost the same, while the $V_1$, $V_2$ and $V_3$ have a nearly gaussian shape. At later redshifts, the center of the MFs shift to larger values of $\delta T_{\rm b}$ and a tail extends to the left, which indicates the non-gaussianity is increasing.

Finally, we consider the reionization stage. The MFs at this stage are shown in Fig.~\ref{evolution1}. The $\rm V_0$ decreases sharply at $\delta T_{\rm b} = 0$ because of the presence of ionized bubbles where $\delta T_{\rm b} = 0$. However, as we can see, the rapid drop is relaxed due to the smoothing, especially when the bubble sizes are smaller than or comparable to the smoothing scale. This feature is typical of the $V_0$ during the late phase of reionization. On the other hand, the peaks in the $V_1$, $V_2$ and $V_3$ shift toward low-temperature direction as reionization proceeds. This is because reionization starts from high-density regions, where the brightness temperature is also high, so that highest-$\delta T_{\rm b}$ regions are abruptly transformed into lowest-$\delta T_{\rm b}$ regions. This drastic change in the distribution of brightness temperature leads to the highly-nongaussian shape of the MFs.

Let us note on the behavior of the $V_3$ at the reionization stage. A peak can be seen around $\delta T_{\rm b} = 0$ at $z = 8.62$, but it disappears and the $V_3$ becomes negative at later time. This implies that ionized regions are isolated at $z = 8.62$ and have merged by $z = 8.06$. This interpretation is supported by the fact that the average neutral fraction decrease from 0.62 to 0.38 during this epoch.

The evolution of MFs before reionization has not been investigated so far and previous works have not studied the evolution of the $V_0$, $V_1$ and $V_2$ even during the EoR. Our results of $V_3$ at reionization stage are consistent with those of previous works \citep{2008ApJ...675....8L,2011MNRAS.413.1353F}. However, the resolution and volume of their simulation maps are smaller than ours. Therefore, our results are complementary to other works. We note that the evolution of the MFs is based on a specific choice of reionization model with a plausible reionization and heating history.

\subsection{Effect of changing $T_{\rm vir}^{\rm min}$ and $\zeta$}

\begin{figure*}
 \centering
  \includegraphics[width=10cm,clip]{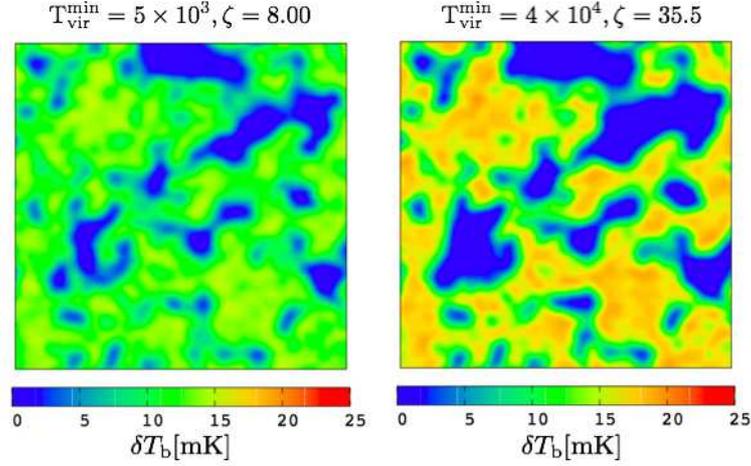}
   \caption{Maps of the brightness temperature at $z$=8.60 with different combinations of $T_{\rm vir}^{\rm min}$ and $\zeta$. } 
 \label{mapTvir} %
\end{figure*}

\begin{figure*}
 \centering
 \includegraphics[width=11cm]{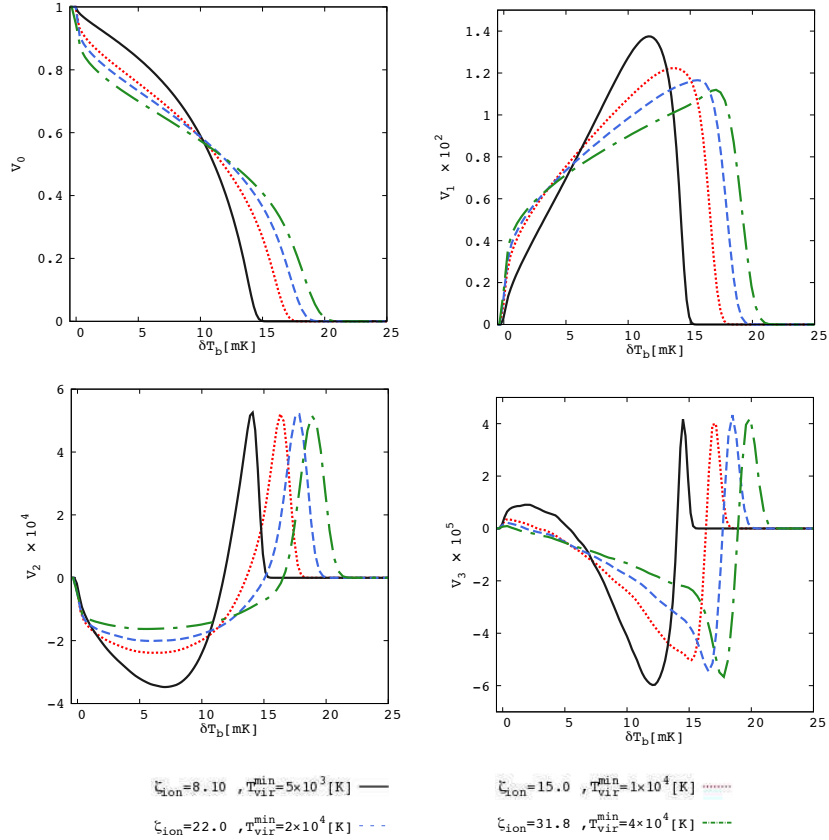}
  \caption{MFs with different combinations of $T_{\rm vir}^{\rm min}$ and $\zeta$. All models have the same neutral fraction $x_{\rm HI} \approx 0.511$ at $z = 8.60$. Here the spin temperature fluctuations are ignored.} 
 \label{Tvir} %
\end{figure*}

The distribution of ionized bubbles and topological structure of brightness temperature are sensitive to the value of $T_{\rm vir}^{\rm min}$ and $\zeta$. In this section, we see the dependence of the MFs on these parameters. The dependence of the MFs on ionizing source properties has been discussed in previous works. For example, \cite{2011MNRAS.413.1353F} has focused on the $V_3$ for the ionized fraction field rather than the $\delta T_{\rm b}$ field. On the other hand, \cite{2014JKAS...47...49H} has explored 2D genus of the $\delta T_{\rm b}$ field.

In Fig.~\ref{mapTvir}, we show $\delta T_{\rm b}$ maps with different combinations of $T_{\rm vir}^{\rm min}$ and $\zeta$, fixing the average neutral fraction to $0.51$ at $z = 8.60$. For the two simulations, we used the same initial condition of matter density fluctuations. As we can see, a lower $T_{\rm vir}^{\rm min}$ results in a larger number of ionizing sources and then ionized bubbles. On the other hand, some of the ionized bubbles are connected for the high-$T_{\rm vir}^{\rm min}$ model because the ionizing efficiency is higher for this model, while the number of ionizing sources is smaller.

In Fig.~\ref{Tvir}, we compare the MFs of four models with a different combination of $T_{\rm vir}^{\rm min}$ and $\zeta$, again fixing the average neutral fraction to $0.51$ at $z = 8.60$. Because ionizing sources exist only in densest regions for the model with a high  $T_{\rm vir}^{\rm min}$, relatively dense regions which do not have ionizing sources remain neutral and have higher values of dTb. Although the neutral fraction is fixed, the decrease in the $\rm V_0$ at $\delta T_{\rm b} = 0~{\rm mK}$ is not the same for the four models. This is due to the effect of smoothing, which is more significant for a model with a low $T_{\rm vir}^{\rm min}$ which leads to a larger number of small bubbles.

A valley of the $V_2$ located at a low $\delta T_{\rm b}$ value represents the property of smoothed ionized bubbles. It becomes deep as $T_{\rm vir}^{\rm min}$ decreases because small bubbles have higher curvature which results in a large absolute value of the $V_2$.

Concerning the $V_3$, the peak around $\delta T_{\rm b}=0$ becomes noticeable as $T_{\rm vir}^{\rm min}$ decreases, which implies that isolated ionized bubbles are more abundant for a lower value of $T_{\rm vir}^{\rm min}$.

\section{Summary and Discussion}

In this work, we investigated the topological structure of the 21cm-line brightness temperature field during and before the EoR by using the Minkowski functionals. Firstly, we compared the MFs of our fiducial model with those of the component maps and found that the main component of the topological structure is the neutral fraction during the EoR and the spin temperature before the EoR. Further, the spin temperature can contribute to the structure of $\delta T_{\rm b}$ even though the average spin temperature is three times higher than $T_{\gamma}$.

Secondly, we studied the evolution of the MFs considering three stages. The evolution in the heating stage is explained by expansion of X-ray-heated regions. In the transition stage, small ionized bubbles and disappearance of cold gas change the structure of $\delta T_{\rm b}$. The complicated evolution in the reionization stage is understood as the expansion and merger of ionized bubbles.

Thirdly, the variation of the MFs changing the values of $T_{\rm vir}^{\rm min}$ and $\zeta$ was investigated. We found that these parameters, even if the average neutral fraction is fixed, affect the topological structure of the IGM and we interpreted the variation of the MFs. Therefore, the MFs could be useful for constraining these model parameters.

This kind of analysis will be viable when the imaging of 21cm signal is possible with a high angular resolution. Practically, there are many obstacles for imaging such as galactic and extragalactic foregrounds. We will study expected constraints on the model parameters with the MFs including these observational difficulty in the future work.

\section*{Acknowledgement}
This work is supported by Grant-in-Aid from the Ministry of Education, Culture, Sports, Science and Technology (MEXT) of Japan, No. 25-3015 (HS), No. 24340048 (KT), No. 26610048 (KT), No. 15H05896 (KT), No. 16H05999 (KT), No. 15K05074 (TM) and No. 15H05890 (TM).





%
%


\bsp	
\label{lastpage}
\end{document}